\documentclass[reprint,aps,prd,onecolumn,notitlepage,nofootinbib,showpacs,preprintnumbers,superscriptaddress]{revtex4-2}
\usepackage[T1]{fontenc}
\usepackage[utf8]{inputenc}
\usepackage{bm}
\usepackage{amsmath}
\usepackage{esint}
\usepackage{color}
\usepackage{graphicx}
\PassOptionsToPackage{normalem}{ulem}
\usepackage{ulem}

\usepackage{hyperref}

\begin{document}
	
	\title{Spatially covariant gravity: Degeneracy condition and mimetic formulation}
	
	\author{Xuezheng Zhu}%
	\email[Email: ]{zhuxzh5@mail2.sysu.edu.cn}
	\affiliation{%
		School of Physics and Astronomy, Sun Yat-sen University, Zhuhai 519088, China}

\begin{abstract} We make a perturbative analysis of spatially covariant gravity only respecting spatial symmetries, of which the Lagrangian includes the dynamic lapse function and the coupling term of spatial curvature and extrinsic curvature. We show that the theory  propagates 2 scalar degrees of freedom, as long as the kinetic terms are degenerate, it propagates a single scalar mode at the linear order. Then we calculate the scalar and tensor power spectra. Finally, we study cosmological perturbations in mimetic gravity in the presence of spatially covariant gravity. The original spatially covariant theory propagates 2 scalar degrees of freedom, and  it will propagate only one scalar degree of freedom after adding the mimetic constraint.
\end{abstract}




\maketitle

\section{Introduction}
The  scalar-tensor theories that include higher derivatives of the scalar field(s) as well as novel couplings between scalar field(s) and the gravity  has been extended significantly in the past decade.
The representative achievements are the $k$-essence\cite{Armendariz-Picon:1999hyi,Chiba:1999ka} and the Horndeski/Galileon theory \cite{Horndeski:1974wa,Deffayet:2011gz,Kobayashi:2011nu}.
The Horndeski theory \cite{Horndeski:1974wa} that beyond the General Relativity is regarded as the most general scalar-tensor theory with a single scalar field, in which both the Lagrangian and the equations of motion are up to the second order in derivatives. Modified gravity theory generalize Horndeski theory has been greatly developed  \cite{Gleyzes:2014dya,Gleyzes:2014qga,Crisostomi:2016czh}, as well as more general higher-order derivative theories with degeneracies  \cite{Gleyzes:2014qga,Gleyzes:2014dya,Crisostomi:2016tcp,Langlois:2015cwa,Crisostomi:2016czh,BenAchour:2016cay,Ezquiaga:2016nqo,Motohashi:2016ftl,Deffayet:2015qwa}, which guarantee the absence of the Ostrogradsky ghost.
See Refs. \cite{Langlois:2018dxi,Kobayashi:2019hrl,Quiros:2019ktw,Amendola:2019laa} for recent reviews on this progress.

   A theory of gravity with spatial covariance has also been developed, compared with general relativity, only covariance with respect to spatial coordinate is retained. It is natural to explore such spatially covariant gravity theories how to extend the scope of scalar-tensor theories. A general frame work for spatially covariant gravity was  developed in \cite{Gao:2014soa,Gao:2014fra}. The Lagrangian is composed of polynomials of the extrinsic curvature $K_{ij}$ — which encodes the velocity of spatial metric $\dot{h_{ij}}$  and generally functions of the lapse function $N$, spatial curvature $R_{ij}$ as well as their spatial derivatives. An important ingredient, however, was omitted in the framework in \cite{Gao:2014soa}, that is the velocity of the lapse function $\dot{N}$. Just like there is a time derivative of the spatial metric $h_{ij}$, we believe the velocity of the lapse function $\dot{N}$ can exist,  shown in \cite{Gao:2018znj,Gao:2019lpz}. Contrary to the theories in Ref. \cite{Gao:2014soa}, generally including $\dot{N}$ in the Lagrangian will introduce an extra scalar-type degree of freedom.

For the spatially covariant gravity with a dynamic lapse function, we find that this theory will propagate two scalar degrees of freedom, so you can treat it as  two fields \cite{Zhu:2022caa}. The purpose of this paper is thus to provide an alternative derivation  in order to evade the unwanted mode.
 we consider linear perturbations around a Friedmann-Robertson-Walker background in order to check under which condition(s) there is only a single scalar mode that propagates.
  We study the degenerate condition of the two-field model and  calculate the scalar spectrum of the single field after degeneration and the tensor spectrum.

Mimetic matter was introduced by Chamseddine and Mukhanov in \cite{Chamseddine:2013kea} to mimic cold dark matter. And it is a modification of General Relativity (GR) leading to a
scalar-tensor type theory, has received considerable attention in cosmology, and has now been extended to inflation, dark energy. The  theory possesses  some interesting features, for example, it was shown that the  theory  contains an extra scalar mode. Mimetic theories have also been studied in \cite{Deruelle:2014zza,Arroja:2015wpa,Arroja:2015yvd,Chamseddine:2014vna,Nojiri:2017ygt,Dutta:2017fjw,Yuan:2015tta,Solomon:2019qgf,Shen:2019nyp,Ganz:2019vre,Ganz:2018vzg,Astashenok:2015haa,Nojiri:2016vhu}. (See \cite{Sebastiani:2016ras} for a
review on mimetic gravity).
Then we study the power spectrum  after adding mimetic constraint to the spatial covariance theory. In unitary gauge, the equation for $\lambda$ is no longer  an equation of motion but leads to a constraint that fixes the lapse to $N = 1$. So we can start by setting $N = 1$ in the action, shown in \cite{Langlois:2018jdg}.

This article  is organized as follows. In sec. \ref{sec:Perturbative} we study field transformation and degeneracy condition of the two fields. In  sec. \ref{sec:tensor specture} we calculate  tensor spectrum. In sec. \ref{sec:gradient instability} and sec. \ref{sec:mimetic gravity} we introduce the mimetic theory and the perturbation of the mimetic spatial covariance theory. In Sec. \ref{sec:conclusion} we summarize our analysis.

\section{ field transformation and degeneracy  condition}\label{sec:Perturbative}
Under the unitary gauge $t=\varphi$, let\textquoteright s
consider a Lagrangian
\begin{align}
\label{action}
\mathcal{L} =&a_{1}K+a_{2}F+b_{1}K_{ij}K^{ij}+b_{2}K^{2}+c_{1}KF\\\nonumber&+c_{2}F^{2}+d_{1}+d_{2}R+a_{3} R^{ij}K_{ij}+d_{3} R^{ij}R_{ij},
\end{align}
compared with the action in \cite{Gao:2019lpz,Zhu:2022caa}, the last two terms are added. Then the action includes the coupling term of spatial curvature and extrinsic curvature.
Where coefficients $a_{i}$ , $b_{i}$ and $d_{i}$ are general functions of
$t$, $N$. $K_{ij}$ is extrinsic curvature and $N$ is lapse function which
can be expressed as

\begin{equation}
\mathcal{L}_{\vec{n}}N=\frac{1}{N}\left(\dot{N}-\mathcal{L}_{\overrightarrow{N}}N\right),
\end{equation}

\begin{equation}
K_{ij}=\frac{1}{2}\mathcal{L}_{\vec{n}}h_{ij}=\frac{1}{2N}\left(\dot{h_{ij}}-\mathcal{L}_{\overrightarrow{N}}h_{ij}\right).
\end{equation}
We can express spacetime metric in term of the ADM
variables

\begin{equation}
\mathrm{d}s^{2}=-N^{2}\mathrm{d}t^{2}+h_{ij}\left(\mathrm{d}x^{i}+N_{i}\mathrm{d}t\right)\left(\mathrm{d}x^{j}+N^{j}\mathrm{d}t\right),
\end{equation}
we can separate the ADM variables to homogeneous
part and perturbations as follows

\begin{align}
 & N=e^{A},\\
 & N_{i}=a B_{i},\\
 & h_{ij}=a^{2}\left(e^{H}\right)_{ij}=a^{2}\left(\delta_{ij}+H_{ij}+1/2H_{i}^{k}H_{kj}+..\right),
\end{align}
as usual, we further decompose $B_{i}$ and $H_{ij}$ into irreducible
parts as

\begin{equation}
B_{i}=\partial_{i}B+S_{i},
\end{equation}
\begin{equation}
H_{ij}=2\zeta\delta_{ij}+\left(\partial_{i}\partial_{j}-\frac{1}{3}\delta_{ij}\partial^{2}\right)E+\partial_{(i}F_{j})+r_{ij},
\end{equation}
with $\partial^{2}$=$\delta^{ij}\partial_{i}$$\partial_{j}$
and $\partial^{i}$$S_{i}$$ $=$\partial^{i}$$F_{i}$$ $=$\partial^{i}$$\gamma_{ij}$=$\gamma_{i}^{i}$=0, and
thus $\zeta$ is identified as the scalar mode and $\gamma_{ij}$
the tensor modes. In order to simplify the calculations, we work in
the gauge with $E=F_{i}=0$.
The quadratic-order action for the  modes in (\ref{action}) takes the general form

\begin{align}
\label{quadratic action}
S_{2}^{\mathrm{S}}\left[\zeta,A,B\right]=\int\mathrm{d}t\mathrm{d}^{3}x\,a^{3}\left(L_{2}^{(1)}+L_{2}^{(2)}\right),
\end{align}
which are terms relevant to counting the number of degrees
of freedom, with
\begin{align}
\label{action L1}
L_{2}^{(1)} & =  \mathcal{C}_{\dot{\zeta}^{2}}\dot{\zeta}^{2}+\mathcal{C}_{\dot{\zeta}\dot{A}}\dot{\zeta}\dot{A}+\mathcal{C}_{\dot{A}^{2}}\dot{A}^{2}\nonumber\\
 & -\mathcal{C}_{\dot{\zeta}B}\dot{\zeta}\frac{\partial^{2}B}{a}-\mathcal{C}_{\dot{A}B}\dot{A}\frac{\partial^{2}B}{a}+\mathcal{C}_{B^{2}}\frac{(\partial^{2}B)^{2}}{a^{2}}+\mathcal{C}_{\zeta B}\frac{1}{a^{3}}\partial^{2}B\partial^{2}\zeta,
\end{align}
terms which are irrelevant to counting the number of degrees
of freedom are

\begin{align}
\label{action L2}
L_{2}^{(2)}  =
\mathcal{C}_{\zeta^{2}}\zeta^{2}+\mathcal{C}_{\dot{\zeta}A}\dot{\zeta}A+\mathcal{C}_{\zeta A}\zeta A+\mathcal{C}_{A^{2}}A^{2}-\mathcal{C}_{AB}A\frac{\partial^{2}B}{a},
\end{align}
the corresponding coefficients are shown in Appendix (\ref{app:coeff}).

The equation of motion for $B$ is
\begin{align}
\label{motion equation B}
\mathcal{C}_{\dot{\zeta}B}\dot{\zeta}+\mathcal{C}_{\dot{A}B}\dot{A}+\mathcal{C}_{AB}A-2\mathcal{C}_{B^{2}}\frac{\partial^{2}B}{a}-\mathcal{C}_{\zeta B}\frac{1}{a^{2}}\partial^{2}\zeta=0,
\end{align}
if $\mathcal{C}_{B^{2}} \neq 0$, from (\ref{motion equation B}) we obtain

\begin{align}
\label{motion equation B_1}
\frac{\partial^{2}B}{a}=\frac{1}{2\mathcal{C}_{B^{2}}}\left(\mathcal{C}_{\dot{\zeta}B}\dot{\zeta}+\mathcal{C}_{\dot{A}B}\dot{A}+\mathcal{C}_{AB}A-\mathcal{C}_{\zeta B}\frac{1}{a^{2}}\partial^{2}\zeta\right).
\end{align}
Substituting (\ref{motion equation B_1}) into (\ref{quadratic action}), we obtain

\begin{align}
S_{2}^{\mathrm{S}}\left[\zeta,A\right] \equiv & \int\mathrm{d}t\frac{\mathrm{d}^{3}x}{\left(2\pi\right)^{3}}a^{3}\,\Big[\mathcal{D}_{\dot{\zeta}^{2}}\dot{\zeta}^{2}+\mathcal{D}_{\dot{\zeta}\dot{A}}\dot{\zeta}\dot{A}+\mathcal{D}_{\dot{A}^{2}}\dot{A}^{2}\nonumber\\
  & +\mathcal{G}_{\zeta^{2}}\zeta^{2}+\mathcal{D}_{\dot{\zeta}A}\dot{\zeta}A+\mathcal{G}_{\zeta A}\zeta A+\mathcal{D}_{A^{2}}A^{2}\Big],\label{S2_{z}eta_{A}}
\end{align}
for the quadratic action (\ref{S2_{z}eta_{A}}), there are  additional terms $(\mathcal{C}_{\dot{A}B}\dot{A}\partial^{2}\zeta+\mathcal{C}_{\dot{\zeta}B}\dot{\zeta}\partial^{2}\zeta)\frac{\mathcal{C}_{\zeta B}}{2\mathcal{C}_{B^{2}}a^{2}}$,
the appearance of these two terms is due to the newly added terms in (\ref{action}).
In which the new coefficients are given by

\begin{align}\mathcal{D}_{\dot{\zeta}^{2}} & = \mathcal{C}_{\dot{\zeta}^{2}}-\frac{1}{4}\frac{\mathcal{C}_{\dot{\zeta}B}^{2}}{\mathcal{C}_{B^{2}}},\label{calD_{z}d2}\\
\mathcal{D}_{\dot{\zeta}\dot{A}} & = \mathcal{C}_{\dot{\zeta}\dot{A}}-\frac{1}{2}\frac{\mathcal{C}_{\dot{\zeta}B}\mathcal{C}_{\dot{A}B}}{\mathcal{C}_{B^{2}}},\\
\mathcal{D}_{\dot{A}^{2}} & =  \mathcal{C}_{\dot{A}^{2}}-\frac{1}{4}\frac{\mathcal{C}_{\dot{A}B}^{2}}{\mathcal{C}_{B^{2}}},\label{calD_{A}d2}
\end{align}

\begin{equation}	\mathcal{G}_{\zeta^{2}}=-2d_{2}\frac{\partial^{2}}{a^{2}}+3d_{3}\frac{\partial^{2}}{a^{2}}+\frac{\mathcal{C}_{\zeta B}^{2}}{4\mathcal{C}_{B^{2}}}(\frac{\partial^{2}}{a^{2}})^{2},
\end{equation}

\begin{equation}
	\mathcal{G}_{\zeta A} = -4\left(d_{2}+\frac{\partial d_{2}}{\partial N}\right)\frac{\partial^{2}}{a^{2}}+a_{3}^{\prime}H\frac{\partial^{2}}{a^{2}}+\frac{\mathcal{C}_{AB}\mathcal{C}_{\zeta B}}{2\mathcal{C}_{B^{2}}}\frac{\partial^{2}}{a^{2}},
	\end{equation}

\begin{align}\mathcal{D}_{\dot{\zeta}A} & =  \mathcal{C}_{\dot{\zeta}A}-\frac{1}{2}\frac{\mathcal{C}_{\dot{\zeta}B}\mathcal{C}_{AB}}{\mathcal{C}_{B^{2}}},\\
\mathcal{D}_{A^{2}} & =  \mathcal{C}_{A^{2}}-\frac{1}{4}\frac{\mathcal{C}_{AB}^{2}}{\mathcal{C}_{B^{2}}}+\frac{1}{4}\frac{1}{a^{3}}\partial_{t}\left(a^{3}\frac{\mathcal{C}_{AB}\mathcal{C}_{\dot{A}B}}{\mathcal{C}_{B^{2}}}\right).
\end{align}

If $A$ field is an auxiliary field, the coefficient of the terms containing the derivative of $A$ should  be 0.
We introduce a variable transformation, see Appendix (\ref{app:degradation})
\begin{align}
\label{transformation}
\tilde{\zeta}:=\zeta+\frac{c_{1}}{2\left(b_{1}+3b_{2}\right)}A,
\end{align}
after the transformation, terms like $\dot{\tilde{\zeta}}\dot{A}$ will disappear, but may containing the term  $\dot{A}^{2}$ and $\dot{A}\partial^{2}\tilde{\zeta}$. We find the coefficient of the term $\dot{A}\partial^{2}\tilde{\zeta}$ is proportional to $-\mathcal{C}_{\dot{A}B}+\frac{c_{1}}{2\left(b_{1}+3b_{2}\right)}\mathcal{C}_{\dot{\zeta}B}$.
We find
\begin{equation}
-\mathcal{C}_{\dot{A}B}+\frac{c_{1}}{2\left(b_{1}+3b_{2}\right)}\mathcal{C}_{\dot{\zeta}B}=-c_{1}+c_{1}=0,
\end{equation}
this means the term $\dot{A}\partial^{2}\tilde{\zeta}$ will disappear directly.
In order to make the coefficients of the term ${\dot{A}}^2$ equal to 0 after field transformation, the degeneracy condition is
\begin{equation}
\label{degeration condition}
4b_1c_2-3{c_1}^2+12b_2c_2=0,
\end{equation}
 which is given in Appendix (\ref{app:degradation})
, the three terms form a perfect  square as
\begin{align}
&
\mathcal{D}_{\dot{\zeta}^{2}}\dot{\zeta}^{2}+\mathcal{D}_{\dot{\zeta}\dot{A}}\dot{\zeta}\dot{A}+\mathcal{D}_{\dot{A}^{2}}\dot{A}^{2}\nonumber\\
  & = \mathcal{D}_{\dot{\zeta}^{2}}\left[\dot{\zeta}+\frac{c_{1}}{2(b_{1}+3b_{2})}\dot{A}\right]^{2}.\label{kin_{p}s}
\end{align}
With the degeneracy condition (\ref{degeration condition}), the $A$ field becomes an auxiliary field, leaving only the scalar $\tilde{\zeta}$ degree of freedom. In special cases, the  solution to the single $\tilde {\zeta}$ mode is in the form of plane waves, shown in Appendix \ref{sec:scalar specture}.

Where $\mathcal{C}_{B^{2}} = 0$, if we deal with constraint equation of $B$, we can't get the relation (\ref{motion equation B_1}).
For the quadratic action (\ref{quadratic action}), we do variable transformation first,
\begin{equation}
\label{transformation1}
\tilde{\zeta}:=\zeta+\frac{c_{1}}{2\left(b_{1}+3b_{2}\right)}A,
\end{equation}
we  find that the coefficient of term $\dot{A}\frac{\partial^{2}B}{a}$ is 0,
\begin{equation}
\label{AB}
-\mathcal{C}_{\dot{A}B}\dot{A}\frac{\partial^{2}B}{a}+\frac{c_{1}}{2\left(b_{1}+3b_{2}\right)}\mathcal{C}_{\dot{\zeta}B}\dot{A}\frac{\partial^{2}B}{a}=0.
\end{equation}
The second term in (\ref{AB}) comes from the field transformation of the term $-\mathcal{C}_{\dot{\zeta}B}\dot{\zeta}\frac{\partial^{2}B}{a}$.
 Using the variable transformation (\ref{transformation1}) and degeneracy condition in (\ref{degeration condition}), we obtain the quadratic-order action as
\begin{align}
\mathcal{L}=&\mathcal{C}_{{\zeta}^{2}}\tilde{\zeta}^{2}+\mathcal{C}_{\dot{\zeta}A}\dot{\tilde{\zeta}}A+\mathcal{U}_{\zeta A}\tilde{\zeta} A+\mathcal{U}_{A^{2}}A^{2}-\mathcal{C}_{AB}A\frac{\partial^{2}B}{a}\nonumber\\&+\mathcal{C}_{\zeta B}\frac{1}{a^{3}}\partial^{2}B\partial^{2}\tilde{\zeta}+\mathcal{C}_{\zeta B}\frac{1}{a^{3}}\partial^{2}B\partial^{2}A+\mathcal{C}_{\dot{\zeta}^{2}}\dot{\tilde{\zeta}}^{2}-\mathcal{C}_{\dot{\zeta}B}\dot{\tilde{\zeta}}\frac{\partial^{2}B}{a},
\end{align}
 in this way, $A$ and $B$ are both auxiliary variable.
We can eliminate $A$ and $B$ ( we first get the equation of motion for $A$ instead of $B$).
So after doing variable transformation (\ref{transformation1}) and combing the degeneracy condition (\ref{degeration condition}), we obtain a theory that only propagates one $\tilde{\zeta}$ scalar degree of freedom.

\section{ tensor spectrum}\label{sec:tensor specture}
 For the Lagrangian in (\ref{action}), the quadratic action for the tensor modes $\gamma_{ij}$ is
\begin{equation}
\label{quadratic action_2}
S_{2}^{\gamma}=\int\!\mathrm{d}t\frac{\mathrm{d}^{3}k}{\left(2\pi\right)^{3}}a^{3}\left(\mathcal{G}_{\gamma}\left(\partial_{t}\gamma_{ij}\right)^{2}-\mathcal{W}_{\gamma}\frac{k^{2}}{a^{2}}\gamma_{ij}^{2}\right),
\end{equation}
where
\begin{eqnarray}
\mathcal{G}_{\gamma} & = & \frac{1}{4}b_{1},\label{Gg}\\
\mathcal{W}_{\gamma} & = & \mathcal{W}_{\gamma}^{(0)}+\mathcal{W}_{\gamma}^{(1)}\frac{k^{2}}{a^{2}}, \label{Wg}
\end{eqnarray}
with
\begin{eqnarray}
\mathcal{W}_{\gamma}^{(0)} & = & \frac{1}{4}\left(d_{2}+a_{3}H\right)\nonumber \\
&  & +\frac{1}{8a}\frac{\mathrm{d}}{\mathrm{d}t}\left( aa_{3}\right) ,\label{Wg0}\\
\mathcal{W}_{\gamma}^{(1)} & = & -\frac{1}{4}d_{3}.\label{Wg1}
\end{eqnarray}

Let us calculate the power spectrum of the tensor mode.
By introducing the conformal time $\tau$ through
        \begin{equation}
       \mathrm{d}t=a\,\mathrm{d}\tau, \label{taudef}
        \end{equation}
the quadratic action (\ref{quadratic action_2}) can be rewritten as
        \begin{equation}
       S_{2}^{\gamma}=\int\!\mathrm{d}\tau\frac{\mathrm{d}^{3}k}{\left(2\pi\right)^{3}}\frac{1}{2}z^{2}\left[\left(\partial_{\tau}\gamma_{ij}\right)^{2}-\left(1+\frac{k^{2}}{a^{2}M^{2}}\right)c_{\gamma}^{2}k^{2}\gamma_{ij}^{2}\right],
                \label{Sgamma2}
        \end{equation}
where we have defined
        \begin{equation}
        z^{2}=2a^{2}\mathcal{G}_{\gamma}, \qquad c_{\gamma}^{2}=\frac{\mathcal{W}_{\gamma}^{(0)}}{\mathcal{G}_{\gamma}},\qquad M^{2}=\frac{\mathcal{W}_{\gamma}^{(0)}}{\mathcal{W}_{\gamma}^{(1)}}. \label{zcM}
        \end{equation}

In order to canonically quantize the system, we define
        \begin{equation}
        \hat{\gamma}_{ij}\equiv\frac{1}{z}\sum_{s=\pm2}\left[u(\tau,\bm{k})e_{ij}^{(s)}(\hat{\bm{k}})\hat{a}_{s}(\bm{k})+u^{\ast}\left(\tau,-\bm{k}\right)e_{ij}^{(s)\ast}(-\hat{\bm{k}})\hat{a}_{s}^{\dagger}(-\bm{k})\right],
        \end{equation}
 $\hat{a}(\bm{k})$ and $\hat{a}^{\dagger}(\bm{k})$ are the annihilation and creation operators with the commutation relation
        \begin{equation}
        \left[\hat{a}_{s}(\bm{k}),\hat{a}_{s'}^{\dagger}(\bm{k}')\right]=\left(2\pi\right)^{3}\delta_{ss'}\delta^{3}\left(\bm{k}-\bm{k}'\right),
        \end{equation}
$e_{ij}^{(s)}(\hat{\bm{k}})$ is the polarization tensor with the helicity states $s=\pm2$, satisfying
        \begin{equation}
        \sum_{i}e_{ii}^{(s)}(\hat{\bm{k}})=\sum_{i}k^{i}e_{ij}^{(s)}(\hat{\bm{k}})=0, \qquad e_{ij}^{(s)\ast}(\hat{\bm{k}})=e_{ij}^{(-s)}(\hat{\bm{k}})=e_{ij}^{(s)}(-\hat{\bm{k}}).
        \end{equation}
After using the normalization conditions
\begin{equation}
\sum_{i,j}e_{ij}^{(s)}(\hat{\bm{k}})e_{ij}^{(s')\ast}(\hat{\bm{k}})=\delta^{ss'},
\end{equation}
 the mode function $u(\tau,\bm{k})$ satisfies the equation of motion
\begin{equation}
\label{tensor motion equation}
        \partial_{\tau}^{2}u(\tau,\bm{k})+\left[\left(1+\frac{k^{2}}{a^{2}M^{2}}\right)c_{\gamma}^{2}k^{2}-\frac{\partial_{\tau}^{2}z}{z}\right]u(\tau,\bm{k})=0.
\end{equation}
Then we define
\begin{equation}
\xi=\frac{H}{c_{\gamma}M},
\end{equation}
mode solutions with this kind of modified dispersion relation containing $k^4$ terms has been studied in \cite{Fujita:2015ymn,Gao:2009ht,HosseiniMansoori:2020mxj}.
The two-point function of $\hat{\gamma}_{ij}$ can now be computed as
        \begin{equation}
        \left\langle \hat{\gamma}_{ij}(\bm{k})\hat{\gamma}_{i'j'}(\bm{k}')\right\rangle =\left(2\pi\right)^{3}\delta^{3}\left(\bm{k}+\bm{k}'\right)\mathcal{P}_{ij,i'j'}(\bm{k}),
        \end{equation}
with
\begin{equation}
\mathcal{P}_{ij,i'j'}=\frac{1}{z^{2}}\left\vert u(\tau,k)\right\vert^{2}\sum_{s=\pm2}e_{ij}^{(s)}(\hat{\bm{k}})e_{i'j'}^{(s)\ast}(\hat{\bm{k}}).
\end{equation}
The total power spectrum of the gravitational waves is given by
        \begin{equation}
        \mathcal{P}_{\gamma}(k)\equiv\frac{k^{3}}{2\pi^{2}}\mathcal{P}_{ij,ij}=\frac{k^{3}}{\pi^{2}}\frac{1}{z^{2}}\left\vert u(\tau,k)\right\vert^{2}. \label{psgen}
        \end{equation}

Now our task is to solve for the mode function. In the case where  $c_{\gamma}$, $M$ and $H$ are constant, while $z$ is proportional to $1/\tau$.
It is convenient to introduce a new evolution parameter $x=-c_{\gamma}\tau k$.
The solution of the equation (\ref{tensor motion equation}) is
\begin{align}u=x^{2}e^{\frac{1}{2}i\xi x^{2}}\left[C_{1}U\left(\mu,\frac{5}{2},-i\xi x^{2}\right)+C_{2}L_{-\mu}^{\frac{3}{2}}\left(-i\xi x^{2}\right)\right],\end{align}
with
\begin{align}\mu=\frac{5}{4}-\frac{i}{4\xi
},\end{align}
we must have $C_{2}=0$, $C_{1}$ is fixed by the Wronskian normalization, we get
\begin{align}C_{1}=\frac{1}{\sqrt{2\xi c_{\gamma}k}}\left(-i\xi\right)^{\mu}.\end{align}
Then the mode solution is written as
\begin{align}
u=\frac{\left(-i \xi \right)^{\mu}}{\sqrt{2 \xi c_{\gamma}k}}x^{2}e^{\frac{1}{2}i \xi x^{2}}U\left(\mu,\frac{5}{2},-i \xi x^{2}\right),\label{ufin}
\end{align}
in large scales when $x\rightarrow 0$,
\begin{align}U\left(\mu,\frac{5}{2},-i \xi x^{2}\right)\rightarrow x^{-3}\frac{\Gamma(\frac{3}{2})(-i \xi)^{-\frac{3}{2}}}{\Gamma(\mu)}.\end{align}
\section{mimetic gravity}\label{sec:gradient instability}
Under the unitary gauge $t=\varphi$, let\textquoteright s
consider a Lagrangian in \cite{Gao:2019lpz}

\begin{align}
\label{action 1}
\mathcal{L} =&a_{1}K+a_{2}F+b_{1}K_{ij}K^{ij}+b_{2}K^{2}+c_{1}KF\\\nonumber&+c_{2}F^{2}+d_{1}+d_{2}R,
\end{align}
where coefficients $a_{i}$ and $b_{i}$ are general functions of
$t$, $N$. $K_{ij}$ is extrinsic curvature and $N$ is lapse function which
can be expressed as

\begin{equation}
\mathcal{L}_{\vec{n}}N=\frac{1}{N}\left(\dot{N}-\mathcal{L}_{\overrightarrow{N}}N\right),
\end{equation}

\begin{equation}
K_{ij}=\frac{1}{2}\mathcal{L}_{\vec{n}}h_{ij}=\frac{1}{2N}\left(\dot{h_{ij}}-\mathcal{L}_{\overrightarrow{N}}h_{ij}\right).
\end{equation}

We use the disformal transform in \cite{Firouzjahi:2017txv}
\begin{equation}
g_{uv}=-\tilde{X}\tilde{g_{uv}},
\end{equation}
 this implies the condition
\begin{equation}
X=g^{uv}\varphi_{u}\varphi_{v}=-1\label{N},
\end{equation}
Choosing the unitary gauge $t=\varphi$,  we obtain $g^{00}$= -$\frac{1}{N^{2}}$.  Then the $X$ can be written as

\begin{equation}
X=-\frac{1}{N^{2}}\label{X},
\end{equation}
combining Equation ({\ref{N}) and Equation ({\ref{X}), we get

\begin{equation}
N=1.
\end{equation}
We can use the Lagrange multiplies fomulation.
 The mimetic
action can be written as

\begin{equation}
S^{\prime}=\int d^{4}x\sqrt{-g}\left[\mathcal{L}+\lambda\left(X+1\right)\right],
\end{equation}
the corresponding Lagrangian $\mathcal{L}+\lambda\left(X+1\right)$ takes the form

\begin{align}
\label{action_2}
\mathcal{L}_{1} & =a_{0}K+b_{1}K^{2}+b_{2}K_{ij}K^{ij}+c_{1}K\mathcal{L}_{\vec{n}}N+c_{2}\left(\mathcal{L}_{\vec{n}}N\right)^{2}\\
 & +d_{0}+d_{1}R+d_{2}\nabla_{i}N\nabla^{i}N\mathcal+\lambda\left(1-\frac{1}{N^{2}}\right),\nonumber \\
\nonumber
\end{align}
we can express spacetime metric in term of the ADM
variables

\begin{equation}
\mathrm{d}s^{2}=-N^{2}\mathrm{d}t^{2}+h_{ij}\left(\mathrm{d}x^{i}+N_{i}\mathrm{d}t\right)\left(\mathrm{d}x^{j}+N^{j}\mathrm{d}t\right),
\end{equation}
we can separate the ADM variables to homogeneous
part and perturbations as follows

\begin{align}
 & N_{i}=a B_{i}\\
 & h_{ij}=a^{2}\left(e^{H}\right)_{ij}=a^{2}\left(\delta_{ij}+H_{ij}+1/2H_{i}^{k}H_{kj}+..\right),
\end{align}
as usual, we further decompose $B_{i}$ and $H_{ij}$ into irreducible
parts as

\begin{equation}
B_{i}=\partial_{i}B+S_{i},
\end{equation}
\begin{equation}
H_{ij}=2\zeta\delta_{ij}+\left(\partial_{i}\partial_{j}-\frac{1}{3}\delta_{ij}\partial^{2}\right)E+\partial_{(i}F_{j})+r_{ij},
\end{equation}
with $\partial^{2}$=$\delta^{ij}\partial_{i}$$\partial_{j}$
and $\partial^{i}$$S_{i}$$ $=$\partial^{i}$$F_{i}$$ $=$\partial^{i}$$\gamma_{ij}$=$\gamma_{i}^{i}$=0, and
thus $\zeta$ is identified as the scalar mode and $\gamma_{ij}$
the tensor modes. In order to simplify the calculations, we work in
the gauge with $E=F_{i}=0$,
which completely fixes the gauge freedom.

From the beginning, we can set $N = 1$ in the action, Straighforward expansion of the Lagrangian to the
linear order in perturbation variable, the backgroud equations of motion
are thus

\begin{equation}
\epsilon_{\zeta}=0,
\end{equation}
with

\begin{equation}
\varepsilon_{\zeta}=d_{0}-3\left(b_{2}+3b_{1}\right)H^{2}-2\left(b_{2}+3b_{1}\right)\dot{H}.
\end{equation}

The quadratic action for the scalar modes takes
the following general form

\begin{align}
S_{2}^{S} & =\int dt\frac{d^{3}k}{\left(2\pi\right)^{3}}\left(g_{\zeta\zeta}\left(\partial_{t}\zeta\right)^{2}+w_{\zeta\zeta}\zeta^{2} +w_{BB}B^{2}+g_{B\zeta}B\partial_{t}\zeta\right),
\end{align}
the variation  of $B$ is

\begin{equation}
2\left(b_{1}+3b_{2}\right)\dot{\zeta}-2\left(b_{1}+b_{2}\right)\frac{\partial^{2}B}{a}=0,
\end{equation}
after eliminating auxiliary variable $B$, we get

\begin{equation}
S_{2}^{\mathrm{S}}[\zeta]\equiv\int\mathrm{d}t\frac{\mathrm{d}^{3}x}{\left(2\pi\right)^{3}}a^{3}\,\Big[\mathcal{C}_{\dot{\zeta}^{2}}\dot{\zeta}^{2}-\mathcal{C}_{\zeta^{2}}/a^2(\partial\zeta)^2],{\label{Szeta}}
\end{equation}
with
\begin{equation}
\mathcal{C}_{\dot{\zeta}^{2}}=\frac{2b_{1}\left(b_{1}+3b_{2}\right)}{b_{1}+b_{2}},
\end{equation}

\begin{equation}
\mathcal{C}_{\zeta^{2}}=-2d_{2}.
\end{equation}
For the tensor modes, the quadratic-order action
is
\begin{equation}
	S_{2}^{\mathrm{T}}=\int\mathrm{d}t\frac{\mathrm{d}^{3}k}{\left(2\pi\right)^{3}}\,\frac{1}{4}a^{3}\left(b_{1}\dot{\gamma}_{ij}\dot{\gamma}^{ij}-d_{2}\frac{k^{2}}{a^{2}}\gamma_{ij}\gamma^{ij}\right).
	\end{equation}
Finally, in order to avoid ghost and instabilities, we must impose the condition

\begin{equation}
\frac{2b_{1}\left(b_{1}+3b_{2}\right)}{b_{1}+b_{2}}>0,
\end{equation}

\begin{equation}
b_{1}>0.
\end{equation}

However, we find that the sign requirements of the tensor model and scalar model for the coefficient $d_2$  are exactly the opposite, so gradient instability will exist. The gradient instability can be eliminated by introducing coupling terms such as $R_{ij}K^{ij}$. In addition, these coupling terms will change the dispersion relation of the scalar perturbation.

\section{mimetic gravity and scalar spectrum}\label{sec:mimetic gravity}

We begin with the action in (\ref{action}),
the corresponding Lagrangian $\mathcal{L}+\lambda\left(X+1\right)$ can
thus be written as

\begin{align}
\label{action_2}
\mathcal{L}_{1} & =a_{1}K+a_{2}F+b_{1}K_{ij}K^{ij}+b_{2}K^{2}+c_{1}KF+c_{2}F^{2}+d_{1}\nonumber\\&+d_{2}R+a_{3} R^{ij}K_{ij}+d_{3} R^{ij}R_{ij} +\lambda\left(1-\frac{1}{N^{2}}\right).
\end{align}

From the beginning, we can set $N = 1$ in the action, then there is only one scalar $\zeta$ field but no scalar $A$ field. Straighforward expansion of the action to the
linear order in perturbation variable, the Lagrangian in (\ref{quadratic action}) takes the following general form
\begin{align}
\label{action_without A}
&
\mathcal{L}  =  \mathcal{C}_{\dot{\zeta}^{2}}\dot{\zeta}^{2} -\mathcal{C}_{\dot{\zeta}B}\dot{\zeta}\frac{\partial^{2}B}{a}+\mathcal{C}_{B^{2}}\frac{(\partial^{2}B)^{2}}{a^{2}}+\mathcal{C}_{\zeta^{2}}\zeta^{2}+\mathcal{C}_{\zeta B}\frac{1}{a^{3}}\partial^{2}B\partial^{2}\zeta,
\end{align}
the equation of motion for $B$ is
\begin{align}
\label{motion of B}
\mathcal{C}_{\dot{\zeta}B}\dot{\zeta}-2\mathcal{C}_{B^{2}}\frac{\partial^{2}B}{a}-\mathcal{C}_{\zeta B}\frac{1}{a^{2}}\partial^{2}\zeta=0,
\end{align}
then we obtain
\begin{align}
\label{solution of B}
\frac{\partial^{2}B}{a}=\frac{1}{2\mathcal{C}_{B^{2}}}(\mathcal{C}_{\dot{\zeta}B}\dot{\zeta}-\mathcal{C}_{\zeta B}\frac{1}{a^{2}}\partial^{2}\zeta),
\end{align}
 inserting (\ref{solution of B}) into (\ref{action_without A}) and integrating by parts, we obtain
\begin{equation}
\mathcal{L}=\mathcal{C}_{\dot{\zeta}^{2}}-\frac{C_{\dot{\zeta}B}^{2}}{4\mathcal{C}_{B^{2}}}\dot{\zeta}^{2}+\frac{\mathcal{C}_{\dot{\zeta}B}\mathcal{C}_{\zeta B}}{8C_{B^{2}}}H\frac{1}{a^2}(\partial\zeta)^2+2d_2\frac{1}{a^2}(\partial\zeta)^2+\frac{3}{4\mathcal{C}_{B^{2}}}\mathcal{C}_{\zeta B}^{2}\frac{1}{a^{4}}(\partial^{2}\zeta)^{2}.
\end{equation}
Then the action can be written as

\begin{equation}
S_{2}^{\mathrm{S}}[\zeta]\equiv\int\mathrm{d}t\frac{\mathrm{d}^{3}x}{\left(2\pi\right)^{3}}a^{3}\,\Big[\mathcal{Y}_{\dot{\zeta}^{2}}\dot{\zeta}^{2}+\mathcal{Y}_{\zeta^{2}}/a^2(\partial\zeta)^2+\mathcal{Z}_{\zeta^{2}}/a^4(\partial^{2}\zeta)^{2}],{\label{Szeta}}
\end{equation}
with
\begin{equation}
\mathcal{Y}_{\dot{\zeta}^{2}}=1-\frac{C_{\dot{\zeta}B}^{2}}{4\mathcal{C}_{B^{2}}},
\end{equation}

\begin{equation}
\mathcal{Y}_{\zeta^{2}}=\frac{\mathcal{C}_{\dot{\zeta}B}\mathcal{C}_{\zeta B}}{8C_{B^{2}}}H+2d_2,
\end{equation}

\begin{equation}
\mathcal{Z}_{\zeta^{2}}=\frac{3}{4\mathcal{C}_{B^{2}}}\mathcal{C}_{\zeta B}^{2},
\end{equation}
 we find the dispersion relation including the term $k^4$ which is similar to  the tensor mode in \ref{sec:tensor specture}.

 We can define
        \begin{equation}
        z^{2}=2a^{2}\mathcal{Y}_{\dot{\zeta}^{2}}, \qquad c_{\gamma}^{2}=\frac{-\mathcal{Y}_{\zeta^{2}}}{\mathcal{Y}_{\dot{\zeta}^{2}}},\qquad M^{2}=\frac{\mathcal{Y}_{\zeta^{2}}}{\mathcal{Z}_{\zeta^{2}}}, \label{zcM}
        \end{equation}
 the  solution to the mode function is
\begin{align}
u=\frac{\left(-i \xi \right)^{\mu}}{\sqrt{2 \xi c_{\gamma}k}}x^{2}e^{\frac{1}{2}i \xi x^{2}}U\left(\mu,\frac{5}{2},-i \xi x^{2}\right),\label{ufin}
\end{align}
with
\begin{align}\mu=\frac{5}{4}-\frac{i}{4\xi
},\end{align}
the definition of $\xi$ is as follows
\begin{equation}
\xi=\frac{H}{c_{\gamma}M}.
\end{equation}
The total power spectrum of the $ zeta $field can be expressed as
        \begin{equation}
        \mathcal{P}_{\zeta}(k)=\frac{k^{3}}{2\pi^{2}}\frac{1}{z^{2}}\left\vert u\right\vert^{2}. \label{psgen}
        \end{equation}
\section{Conclusion}\label{sec:conclusion}
When  the spatially covariant gravity contains the  dynamic lapse function $N$, such a kind of theories propagate two scalar-type degrees of freedom. Using the field transformation (\ref{transformation}) and  the degeneracy condition (\ref{degeration condition}), we find the theory will propagate only one $\tilde{\zeta}$ scalar degree of freedom and  the $A$  field becomes an auxiliary field. And the coefficient of the term $\dot{A}\partial^{2}\tilde{\zeta}$ is exactly 0, which does not lead to additional degeneracy condition. Then we calculate the scalar spectrum of the single field after degeneration as well as  the tensor spectrum. For the  tensor mode, its dispersion relation contains $k^4$ terms. Finally we explore the cosmological implications of mimetic models. We study spatially covariant gravity with the coupling term of spatial curvature and extrinsic curvature. After adding the mimetic constraint, the mimetic spatially covariant gravity only propagates a  scalar degree of freedom.
\section*{Acknowledgements}

This work was partly supported by the Natural Science Foundation of China (NSFC) under the grant No. 11975020.

\appendix

\section{Coefficients} \label{app:coeff}

Which are coefficients of the quadratic action, with
	\begin{eqnarray}
	\mathcal{C}_{\dot{\zeta}^{2}} & = & 3(b_{1}+3b_{2}),\\
	\mathcal{C}_{\dot{\zeta}\dot{A}} &= & 3c_{1},\\
	\mathcal{C}_{\dot{A}^{2}} & = & c_{2},\\
	\mathcal{C}_{\dot{\zeta}B} & = & 2(b_{1}+3b_{2}),\\
	\mathcal{C}_{\dot{A}B} & = & c_{1},\\
	\mathcal{C}_{B^{2}} & = & b_{1}+b_{2},
	\end{eqnarray}
\begin{equation}	\mathcal{C}_{\dot{\zeta}A}=-3H\left[2\left(b_{1}+3b_{2}\right)-2\frac{\partial\left(b_{1}+3b_{2}\right)}{\partial N}+3c_{1}\right]+3\left(\frac{\partial a_{1}}{\partial N}-a_{2}\right),
\end{equation}
	
\begin{equation}
\mathcal{C}_{AB}=\frac{1}{3}\mathcal{C}_{\dot{\zeta}A},
\end{equation}
\begin{eqnarray}
	\mathcal{C}_{A^{2}} & = & -\frac{3}{2}\frac{\partial c_{1}}{\partial N}\dot{H}+\frac{1}{2}\left(d_{1}+3\frac{\partial d_{1}}{\partial N}+\frac{\partial^{2}d_{1}}{\partial N^{2}}\right)\nonumber \\
	&  & +\frac{3}{2}H^{2}\left[b_{1}+3b_{2}-\frac{\partial\left(b_{1}+3b_{2}\right)}{\partial N}+\frac{\partial^{2}\left(b_{1}+3b_{2}\right)}{\partial N^{2}}-3\frac{\partial c_{1}}{\partial N}\right] \\
	&  & +\frac{3}{2}H\left(\frac{\partial a_{1}}{\partial N}-a_{2}+\frac{\partial^{2}a_{1}}{\partial N^{2}}-\frac{\partial a_{2}}{\partial N}\right),\nonumber
	\end{eqnarray}
\begin{equation}
	\mathcal{C}_{\zeta^{2}}=-2d_{2}\frac{\partial^{2}}{a^{2}}+3d_{3}\frac{\partial^{2}}{a^{2}},
	\end{equation}
\begin{equation}
	\mathcal{C}_{\zeta B} = 2a_{3},
	\end{equation}
\begin{equation}
	\mathcal{C}_{\zeta A} = -4\left(d_{2}+\frac{\partial d_{2}}{\partial N}\right)\frac{\partial^{2}}{a^{2}}+a_{3}^{\prime}H\frac{\partial^{2}}{a^{2}}.
\end{equation}

\section{Degeneracy condition} \label{app:degradation}
The coefficients for the kinetic terms can be evaluated explicitly to be
	\begin{eqnarray}
    \label{kinetic coefficients}
	\mathcal{D}_{\dot{\zeta}^{2}} & = & \frac{2b_{1}(b_{1}+3b_{2})}{b_{1}+b_{2}},\nonumber\\
	\mathcal{D}_{\dot{\zeta}\dot{A}} & = & \frac{2b_{1}c_{1}}{b_{1}+b_{2}},\\
	\mathcal{D}_{\dot{A}^{2}} & = & c_{2}-\frac{c_{1}^{2}}{4(b_{1}+b_{2})}\nonumber.
	\end{eqnarray}
The kinetic terms can be written as
\begin{align}
\label{kinetic terms}
&
\mathcal{D}_{\dot{\zeta}^{2}}\dot{\zeta}^{2}+\mathcal{D}_{\dot{\zeta}\dot{A}}\dot{\zeta}\dot{A}+\mathcal{D}_{\dot{A}^{2}}\dot{A}^{2}	
\nonumber\\&=\mathcal{D}_{\dot{\zeta}^{2}}\left(\dot{\zeta}+\frac{\mathcal{D}_{\dot{\zeta}\dot{A}}}{2\mathcal{D}_{\dot{\zeta}^{2}}}\dot{A}\right)^2+\mathcal{D}_{\dot{A}^{2}}\dot{A}^{2}
-\mathcal{D}_{\dot{\zeta}^{2}}\frac{\mathcal{D}_{\dot{\zeta}\dot{A}}^{2}}{4\mathcal{D}_{\dot{\zeta}^{2}}^2}\dot{A}^{2},
\end{align}
using coefficients in (\ref{kinetic coefficients}), we obtain
\begin{equation}
\frac{\mathcal{D}_{\dot{\zeta}\dot{A}}}{2\mathcal{D}_{\dot{\zeta}^{2}}}=\frac{c_{1}}{2(b_{1}+3b_{2})},
\end{equation}
for (\ref{kinetic terms}), the degeneracy condition is
\begin{equation}
\label{degenerate condition 1}
\mathcal{D}_{\dot{A}^{2}}
-\mathcal{D}_{\dot{\zeta}^{2}}\frac{\mathcal{D}_{\dot{\zeta}\dot{A}}^{2}}{4\mathcal{D}_{\dot{\zeta}^{2}}^2}=0,
\end{equation}
using coefficients in (\ref{kinetic coefficients}), the relation (\ref{degenerate condition 1}) leads to
\begin{equation}
4b_1c_2-3{c_1}^2+12b_2c_2=0,
\end{equation}
with the degeneracy condition, we obtain
\begin{align}
&
\mathcal{D}_{\dot{\zeta}^{2}}\dot{\zeta}^{2}+\mathcal{D}_{\dot{\zeta}\dot{A}}\dot{\zeta}\dot{A}+\mathcal{D}_{\dot{A}^{2}}\dot{A}^{2}\nonumber\\
  & = \mathcal{D}_{\dot{\zeta}^{2}}\left[\dot{\zeta}+\frac{c_{1}}{2(b_{1}+3b_{2})}\dot{A}\right]^{2}.\label{kin_{p}s}
\end{align}

\section{Scalar spectrum }\label{sec:scalar specture}
 Under which condition the two fields will degenerate into a field had been discussed. However, the action of the $\tilde{\zeta}$ single field  was not written. We will calculate the  form of the single field, and get its power spectrum. We choose a special and simple case where the coefficient $\mathcal{C}_{\zeta B}$ in (\ref{action L1}) is 0 ( where the coefficient $\mathcal{C}_{\zeta B}\neq 0$, the solution of $\tilde{\zeta}$ will be very complicated), and assume the coefficients of quadratic-order terms and $H$ to be constants.

When the coefficient $\mathcal{C}_{B^{2}} \neq 0$, using transformation (\ref{transformation}) and the degeneracy condition (\ref{degeration condition}), we solve the $\tilde{\zeta}$ dynamic field as
\begin{align}
\label{S2_zetah_A1}
S_{2}^{\mathrm{S}}\left[\tilde{\zeta},A\right]  = & \int\mathrm{d}t\frac{\mathrm{d}^{3}k}{\left(2\pi\right)^{3}}a^{3}\,\Big(\mathcal{D}_{\dot{\zeta}^{2}}\dot{\tilde{\zeta}}^{2}+\mathcal{W}_{\zeta^{2}}\tilde{\zeta}^{2}\nonumber\\
& +\mathcal{F}_{\dot{\zeta}A}\dot{\tilde{\zeta}}A+\mathcal{F}_{\zeta A}\tilde{\zeta}A+\mathcal{F}_{A^{2}}A^{2}\Big),
\end{align}
the coefficients in (\ref{S2_zetah_A1}) are
\begin{eqnarray}
\mathcal{F}_{\dot{\zeta}A} & = & \mathcal{D}_{\dot{\zeta}A},\\
\mathcal{F}_{\zeta A} & = & \mathcal{W}_{\ensuremath{\zeta}A}k^{2}-2\mathcal{W}_{\zeta^{2}}k^{2}\sqrt{\frac{\mathcal{D}_{\dot{A}^{2}}}{\mathcal{D}_{\dot{\zeta}^{2}}}},\\
\mathcal{F}_{A^{2}} & = & \mathcal{D}_{A^{2}}+\mathcal{W}_{\zeta^{2}}k^{2}\frac{\mathcal{D}_{\dot{A}^{2}}}{\mathcal{D}_{\dot{\zeta}^{2}}}-\mathcal{W}_{\zeta A}k^{2}\sqrt{\frac{\mathcal{D}_{\dot{A}^{2}}}{\mathcal{D}_{\dot{\zeta}^{2}}}}\nonumber \\
&  &
+
\frac{3}{2}H\mathcal{D}_{\dot{\zeta}A}\sqrt{\frac{\mathcal{D}_{\dot{A}^{2}}}{\mathcal{D}_{\dot{\zeta}^{2}}}},
\end{eqnarray}
with
\begin{equation}
	\mathcal{W}_{\zeta^{2}}=-2d_{2},
	\end{equation}
\begin{equation}
	\mathcal{W}_{\zeta A} = -4\left(d_{2}+\frac{\partial d_{2}}{\partial N}\right).
	\end{equation}
From (\ref{S2_zetah_A1}), the equation of motion for $A$ is
\begin{equation}
\mathcal{F}_{\dot{\zeta}A}\dot{\tilde{\zeta}}+\mathcal{F}_{\zeta A}\tilde{\zeta}+2\mathcal{F}_{A^{2}}A=0,
\end{equation}
from which we solve
\begin{equation}
A=-\frac{\mathcal{F}_{\dot{\zeta}A}}{2\mathcal{F}_{A^{2}}}\dot{\tilde{\zeta}}-\frac{\mathcal{F}_{\zeta A}}{2\mathcal{F}_{A^{2}}}\tilde{\zeta}.\label{A_sol}
\end{equation}
Substituting (\ref{A_sol}) into (\ref{S2_zetah_A1}), we obtain

\begin{align}
S_{2}^{\mathrm{S}}\left[\tilde{\zeta},A\right]  = & \int\mathrm{d}t\frac{\mathrm{d}^{3}k}{\left(2\pi\right)^{3}}a^{3}\,\left(\mathcal{G}{\tilde{\zeta}}^{2}+\mathcal{W}\tilde{\zeta}^{2}\right),\label{S2_{z}etah_{A}}
\end{align}
with
\begin{eqnarray}
\mathcal{G} & \equiv & \mathcal{D}_{\dot{\zeta}^{2}}-\frac{\mathcal{F}_{\dot{\zeta}A}^{2}}{4\mathcal{F}_{A^{2}}},\\
\mathcal{W} & \equiv & \mathcal{W}_{\zeta^{2}}k^{2}-\frac{\mathcal{F}_{\zeta A}^{2}}{4\mathcal{F}_{A^{2}}}+\frac{3}{4}H\frac{\mathcal{F}_{\zeta A}\mathcal{F}_{\dot{\zeta}A}}{\mathcal{F}_{A^{2}}}.
\end{eqnarray}
For terms of order $\mathcal{O}(\frac{1}{k^2})$, which we neglect in the following analysis, we write the motion equation as

\begin{equation}
\label{equation zeta}
S_{2}^{\mathrm{S}}[\zeta]\equiv\int\mathrm{d}t\frac{\mathrm{d}^{3}x}{\left(2\pi\right)^{3}}a^{3}\,\Big[\mathcal{C}_{\dot{\zeta}^{2}}\dot{\tilde{\zeta}}^{2}-\mathcal{C}_{\zeta^{2}}/a^2(\partial\tilde{\zeta})^2],
\end{equation}
 the coefficients  $\mathcal{C}_{\dot{\zeta}^{2}},\mathcal{C}_{\zeta^{2}}$ can be taken as  constant in special cases.
The equation (\ref{equation zeta}) can be written as
 \begin{equation}{\label{S2}}
         S_{2}^{\mathrm{S}}[\zeta]= \int d\eta d^3x a^2\frac{\epsilon_s}{c_s^2}  \left((\tilde{\zeta}^{\prime})^{2} - c_s^2 (\partial\tilde{\zeta})^2\right),
    \end{equation}
  \begin{eqnarray}
        \epsilon_s &\equiv&\mathcal{C}_{\zeta^{2}},\\
        c_s^2 & = & \frac{\mathcal{C}_{\zeta^{2}}}{\mathcal{C}_{\dot{\zeta}^{2}}},\label{cs_def}
  \end{eqnarray}
we obtain
\begin{align}
\tilde{\zeta}_{k}(\tau,k)=\frac{iH}{2\sqrt{\epsilon_{s}c_{s}k^{3}}}\left(1+ic_{s}k\eta\right)e^{-ic_{s}k\eta},
\end{align}
where all quantities are evaluated around the Hubble exit at $\vert c_s k \eta\vert =1$.
The total power spectrum of the scalar is given by
        \begin{equation}
        \mathcal{P}_{\gamma}(k)=\frac{k^{3}}{2\pi^{2}}\left\vert\tilde{\zeta}_{k}(\tau,k) \right\vert^{2}. \label{psgen}
        \end{equation}

\bibliography{bibfile3}
\end{document}